\documentclass[11pt]{article}

 \def\be{\begin{equation}}
 \def\ee{\end{equation}}
 \def\bea{\begin{eqnarray}}
 \def\eea{\end{eqnarray}}

\def\1{\'{\i}}                           
\def\>#1{{\bf #1}}                 
\def\d{{\rm d}}

\def\te{\theta}
\def\rr{\rho}
\def\tes{\phi}

\def\jp{J_+}
\def\jm{J_-}
\def\jj{J_3}

\def\SW{\rm I}

\def\la{\lambda}
\def\pot{{\cal U}}

\begin{document}

 \thispagestyle{empty}

 \ 
 \hfill\

 \begin{center}

 {\LARGE{\bf{Integrable geodesic motion on 3D curved spaces}}} 

 {\LARGE{\bf{from
  non-standard quantum  deformations}}}

\bigskip

 \bigskip

 \end{center}

 \bigskip 
 \medskip

 \begin{center}   {\sc \'Angel~Ballesteros$^{a,}$\footnote{Communication presented
at the $14^{\rm th}$   International Colloquium on ``Integrable Systems" 14--16
June 2005, Prague, Czech Republic.},
  Francisco~J.~Herranz$^a$,  and Orlando Ragnisco$^{b}$ }
 \end{center}

 \begin{center} {\it { 
 ${}^a$Departamento de F\1sica, Universidad de Burgos, Pza.\
 Misael Ba\~nuelos s.n., \\ E-09001 Burgos, Spain }}\\ e-mail:
 angelb@ubu.es, fjherranz@ubu.es
 \end{center}

 \begin{center} {\it { 
 ${}^b$Dipartimento di Fisica,   Universit\`a di Roma Tre and
 Instituto Nazionale di Fisica Nucleare sezione di Roma Tre,  Via Vasca
 Navale 84,  I-00146 Roma, Italy}}\\ e-mail: ragnisco@fis.uniroma3.it
 \end{center}

 \bigskip 
 \medskip

 \begin{abstract}
 The link between 3D
 spaces with
 (in general, non-constant) curvature and quantum deformations is
presented.  It is shown how the non-standard deformation of a $sl(2)$
Poisson coalgebra
 generates a family of integrable Hamiltonians that represent geodesic
 motions on 3D manifolds with  a non-constant curvature that turns out
to be a function of the
 deformation parameter $z$. A different  Hamiltonian defined on the
same deformed coalgebra is also shown to generate a maximally 
 superintegrable geodesic motion on   3D Riemannian   and $(2+1)$D
relativistic spaces whose  sectional curvatures  
  are all constant and equal to $z$. This approach  can be generalized
to arbitrary dimension.
\end{abstract}

 \bigskip\medskip 



 \bigskip\medskip 

 \newpage


\section{Introduction}

Recently, a non-standard deformation of the Poisson $sl(2)$ coalgebra
has been used to obtain a very large family of (super)integrable 
geodesic motions on certain two-dimensional (2D) spaces with (constant
or variable) curvature~\cite{plb}. The spaces with constant curvature
come from the {\em superintegrable}  geodesic motion; these are  the
sphere,  Euclidean,   hyperbolic,  (anti-)de Sitter and Minkowskian
spaces. For all of them the deformation parameter
$z$  coincides exactly with the Gaussian curvature.   In turn, the
spaces with variable curvature  arise from the {\em integrable}
geodesic motion and they  can be considered as deformations of the
abovementioned spaces for which the curvature is a function of both
the deformation parameter
$z$ and  some intrinsic coordinates of the space. Moreover, some known
and  new   (super)integrable  potentials (such as oscillator  and
Kepler-type ones) defined on these curved spaces have also been
obtained~\cite{potJPA} by making use of the   dynamical coalgebra
symmetry~\cite{BR}.

Although the quantum coproduct ensures the existence of the
generalization of this ``dynamical generation" of curvature to
arbitrary dimension, the explicit geometric characterization of the
spaces so obtained is far from being straightforward. The aim of this
contribution is to present a first step in this direction by making
fully explicit the class of 3D curved spaces together with their
associated free Hamiltonians that arise from this
$q$-deformed geodesic dynamics.


\section{Integrable geodesic motion on 2D curved spaces}

Let us briefly recall the 2D results  by considering the non-standard
quantum deformation of
$sl(2)$ written as a Poisson coalgebra $(sl_z(2), \Delta_z)$ with  real
deformation parameter $z$; the  Poisson brackets,  coproduct and 
Casimir are given by~\cite{Deform}
\be  \{\jj,\jp\}=2 \jp \cosh z\jm  , \qquad  
\{\jj,\jm\}=-2\,\frac {\sinh z\jm}{z} ,\qquad   
\{\jm,\jp\}=4 \jj   , 
\label{ba}
\ee
\be  
\begin{array}{l}
\Delta_z(\jm)=  \jm \otimes 1+ 1\otimes \jm ,\\[2pt]
\Delta_z(J_l)=J_l \otimes {\rm e}^{z \jm} + {\rm e}^{-z \jm} \otimes
J_l   ,\quad l=+,3,
\end{array}
\label{bb}
\ee 
\be {\cal C}_z= \frac {\sinh z\jm}{z} \,\jp -\jj^2  . 
\label{bc}
\ee Starting from the   one-particle symplectic realization of
(\ref{ba})   given by
\be  
 \jm^{(1)}=q_1^2 ,\qquad   \jp^{(1)}=
  \frac {\sinh z q_1^2}{z q_1^2}\,   p_1^2 ,\qquad \jj^{(1)}=
\frac {\sinh z q_1^2}{z q_1^2 }\,    q_1 p_1  ,
\label{bdec}
\ee  (under the which ${\cal C}_z^{(1)}=0$), the coproduct (\ref{bb})
provides the following two-particle symplectic realization of
(\ref{ba}):
\be  
\begin{array}{l}
\displaystyle{ \jm^{(2)}=q_1^2+q_2^2,\qquad   \jp^{(2)}=
  \frac {\sinh z q_1^2}{z q_1^2}\,   p_1^2 \,  {\rm e}^{z q_2^2}+
 \frac {\sinh z q_2^2}{z q_2^2} \,    p_2^2 \,{\rm e}^{-z q_1^2}
},\\[10pt]
\displaystyle{   \jj^{(2)}=
\frac {\sinh z q_1^2}{z q_1^2 }\,  q_1 p_1 \,  {\rm e}^{z q_2^2}  +
\frac {\sinh z q_2^2}{z q_2^2 }\,   q_2 p_2 \, {\rm e}^{-z q_1^2} }.
\end{array}
\label{bec}
\ee 
  By substituting (\ref{bec}) in (\ref{bc}) we obtain
 the two-particle Casimir  
\be 
  {\cal C}_z^{(2)} = \frac {\sinh z q_1^2 }{z q_1^2 } \, \frac {\sinh
z q_2^2}{z q_2^2} 
\left({q_1}{p_2} - {q_2}{p_1}\right)^2 {\rm e}^{-z q_1^2}{\rm e}^{z
q_2^2}   ,
\label{bf}
\ee   which Poisson-commutes with the generators (\ref{bec}).  

The coalgebra approach~\cite{BR}  ensures that {\em any} smooth
Hamiltonian function  ${\cal H}_z={\cal
H}_z(\jm^{(2)},\jp^{(2)},\jj^{(2)})$ gives rise to an integrable
system,  for which  ${\cal C}_z^{(2)}$  is the constant of the motion.
In particular,   we find a large family of integrable deformations of
the free motion  of a particle on the 2D Euclidean space  defined by 
\be  
\begin{array}{l} {\cal H}_z=\frac 12 \jp^{(2)}\, f (z\jm^{(2)} ),
\end{array}
\label{ahaa}
\ee 
where $f$ is an arbitrary  smooth function such that $\lim_{z\to
0}f(zJ_-^{(2)})=1$, i.e., $\lim_{z\to 0}{\cal H}_z=\frac
12(p_1^2+p_2^2)$. Two relevant possibilities for ${\cal H}_z$ have
been studied in~\cite{plb}:

\noindent $\bullet$ The {\em integrable} Hamiltonian 
${\cal H}_z^{\rm I}=\frac 12 \jp^{(2)}$, which defines the  geodesic
motion on  a 2D Riemannian space with  metric 
\be
\d s^2_{\rm I}=\frac {2z q_1^2}{\sinh z q_1^2} \, {\rm e}^{-z q_2^2}
\,\d q_1^2   +
 \frac {2 z q_2^2}{\sinh z q_2^2} \, {\rm e}^{z q_1^2}\, \d q_2^2 ,   
\label{cc}
\ee and whose non-constant   Gaussian curvature reads 
$$K=-   z \sinh\left(z(q_1^2+q_2^2) \right) .$$ 

\noindent $\bullet$ The {\em superintegrable} Hamiltonian 
${\cal H}_z^{\rm S}=\frac 12 \jp^{(2)} \,{\rm e}^{ z \jm^{(2)}}$,
which is  a   St\"ackel  system~\cite{Per} so that this is  endowed
with an additional  constant of the motion~\cite{Deform}
\be {\cal I}_z^{(2)}=\frac {\sinh z q_1^2}{2 z q_1^2} \, {\rm e}^{z
q_1^2}  p_1^2  .
\label{bjj}
\ee This Hamiltonian leads to a  Riemannian   metric of constant
curvature  which coincides with the deformation parameter, $K=z$,
namely
\be
\d s^2_{\rm S}=\frac {2z q_1^2}{\sinh z q_1^2} \, {\rm e}^{-z
q_1^2}{\rm e}^{-2 z q_2^2} \,\d q_1^2   +
 \frac {2 z q_2^2}{\sinh z q_2^2} \, {\rm e}^{-z q_2^2} \, \d q_2^2   .
\label{ec}
\ee

A suitable change of coordinates~\cite{plb} (depending on an
additional ``contraction" parameter) have allowed us to derived the 2D
sphere, Euclidean,  hyperbolic, Minkowskian and (anti-)de Sitter spaces
from the metric  (\ref{ec}). Likewise, their ``deformed" counterpart
(understood as spaces with  non-constant curvature) have been deduced
from the integrable metric (\ref{cc}).
  The explicit solution of the geodesic flows for all these spaces has
been studied in~\cite{potJPA}, as well as a method to introduce
(super)integrable potentials on them by adding a potential term of the
form
$\pot(z\jm^{(2)})$; in this way, some known potentials are recovered
(appearing in the classifications~\cite{RS,PogosClass1}) and also  new
ones are obtained.   We  recall that another approach   to
superintegrability on 2D spaces of variable curvature can be found
in~\cite{KKWa,KKWb}.


\section{Integrable geodesic motion on 3D curved spaces}         

In order to perform the generalization of this construction to 3D
spaces, a three particle symplectic realization of the deformed
Poisson algebra (\ref{ba}) has to be obtained  from the 3-sites
coproduct map~\cite{BR}, which is defined as:
\be
\Delta_z^{(3)} =(\Delta_z \otimes \mbox{id})\circ
\Delta_z =(\mbox{id}\otimes \Delta_z )\circ \Delta_z .
\label{3cop}
\ee Hence from (\ref{bb}) and (\ref{bdec}), we find that (\ref{3cop})
reads
\be  
\begin{array}{l}
 \jm^{(3)}=q_1^2+q_2^2+q_3^2\equiv \>q^2 , \\[2pt]
\displaystyle{ \jp^{(3)}=
  \frac {\sinh z q_1^2}{z q_1^2}  p_1^2 {\rm e}^{z q_2^2}{\rm e}^{z
q_3^2} +
 \frac {\sinh z q_2^2}{z q_2^2}  p_2^2   {\rm e}^{-z q_1^2} {\rm e}^{z
q_3^2}
 +\frac {\sinh z q_3^2}{z q_3^2}  p_3^2    {\rm e}^{-z q_1^2} {\rm
e}^{-z q_2^2}   },\\[6pt]
\displaystyle{  \jj^{(3)}=
\frac {\sinh z q_1^2}{z q_1^2 }  q_1 p_1   {\rm e}^{z q_2^2} {\rm
e}^{z q_3^2}+
\frac {\sinh z q_2^2}{z q_2^2 }  q_2 p_2   {\rm e}^{-z q_1^2}{\rm
e}^{z q_3^2} +
\frac {\sinh z q_3^2}{z q_3^2 }  q_3 p_3   {\rm e}^{-z q_1^2}{\rm
e}^{-z q_2^2}  }.
\end{array}
\label{3symp}
\ee By subsituting  these expresions  in  (\ref{bc}) we get the
three-particle Casimir
\bea && {\cal C}_z^{(3)} = \frac {\sinh z q_1^2 }{z q_1^2 } \,
\frac {\sinh z q_2^2}{z q_2^2} 
\left({q_1}{p_2} - {q_2}{p_1}\right)^2 {\rm e}^{-z q_1^2}{\rm e}^{z
q_2^2} {\rm e}^{2 z q_3^2} \nonumber\\  &&\qquad\qquad+
\frac {\sinh z q_1^2 }{z q_1^2 } \,
\frac {\sinh z q_3^2}{z q_3^2} 
\left({q_1}{p_3} - {q_3}{p_1}\right)^2 {\rm e}^{-z q_1^2} {\rm e}^{  z
q_3^2}\label{3cas} \\ &&\qquad\qquad +
\frac {\sinh z q_2^2 }{z q_2^2 } \,
\frac {\sinh z q_3^2}{z q_3^2} 
\left({q_2}{p_3} - {q_3}{p_2}\right)^2 {\rm e}^{-2z q_1^2}{\rm e}^{-z
q_2^2} {\rm e}^{z q_3^2}  ,
\nonumber
\eea  which Poisson-commutes, by construction~\cite{BR}, with the
three-particle generators (\ref{3symp}) and also with the two-particle
Casimir ${\cal C}_z^{(2)}$ (\ref{bf}). Hence the generic Hamiltonian
${\cal H}_z={\cal H}_z(\jm^{(3)},\jp^{(3)},\jj^{(3)})$ determines a
family of integrable systems as the three functionally independent
functions $\{ {\cal H}_z, {\cal C}_z^{(2)}, {\cal C}_z^{(3)}\}$ are   
mutually in involution.


\subsection{Integrable geodesic motion on spaces of non-constant
curvature}

As in the 2D case, we  consider the  {kinetic energy} ${\cal T}^{\rm
I}_z(q_i,\dot q_i)$ coming from the {\em integrable} Hamiltonian 
${\cal H}_z^{\rm I}(q_i,p_i)=\frac 12 \jp^{(3)}$ that can be rewritten
as the free Lagrangian
\be
 2{\cal T}^{\rm I}_z= \frac
 {z q_1^2}{\sinh z q_1^2} \, {\rm e}^{-z q_2^2}{\rm e}^{-z q_3^2} \dot
q_1^2   +
 \frac {z q_2^2}{\sinh z q_2^2} \, {\rm e}^{z q_1^2}{\rm e}^{-z q_3^2}
\dot q_2^2    +
 \frac {z q_3^2}{\sinh z q_3^2} \, {\rm e}^{z q_1^2}{\rm e}^{z q_2^2}
\dot q_3^2  ,
 \label{ca}
\ee 
 which defines a geodesic flow on a 3D
 Riemannian space with a definite positive  metric   given, up to a
constant factor, by
\be
 \d s^2_{\rm I}=\frac {2z q_1^2}{\sinh z
 q_1^2} \, {\rm e}^{-z q_2^2}{\rm e}^{-z q_3^2} \,\d q_1^2   +
  \frac {2 z q_2^2}{\sinh z q_2^2} \, {\rm e}^{z q_1^2}{\rm e}^{-z
q_3^2}\, \d q_2^2   +
  \frac {2 z q_3^2}{\sinh z q_3^2} \, {\rm e}^{z q_1^2}{\rm e}^{z
q_2^2}\, \d q_3^2 .
 \label{ccd}
\ee The sectional curvatures $K_{ij}$ in the planes 12, 13 and 23 turn
out to be
\be  
\begin{array}{l} K_{12}=\frac z4 \,{\rm e}^{-z \>q^2}\bigl( 1+ {\rm
e}^{2 z q_3^2}- 2 {\rm e}^{2z \>q^2}\bigr) ,  \\[2pt] K_{13}=\frac z4
\,{\rm e}^{-z \>q^2}\bigl( 2- {\rm e}^{2 z q_3^2}+ {\rm e}^{2 z
q_2^2}{\rm e}^{2 z q_3^2}- 2 {\rm e}^{2z
\>q^2}\bigr)  , \\[2pt] K_{23}=\frac z4 \,{\rm e}^{-z \>q^2}\bigl( 2- 
{\rm e}^{2 z q_2^2}{\rm e}^{2 z q_3^2}- 2 {\rm e}^{2z
\>q^2}\bigr)  ,    
\end{array}
\label{xa}
\ee while the  scalar curvature $K$  fulfills
\be K=2(K_{12}+K_{13}+K_{23})=-5 z \sinh(z\>q^2) .
\label{xb}
\ee 

The geometric characterization of these spaces   becomes much more
clear if we introduce the new canonical coordinates
$(\rho,\te,\tes)$ and conjugated momenta $(p_\rho,p_\te,p_\tes)$
defined by
\bea  &&
\cosh^2(\la_1\rho)= {\rm e}^{2z \>q^2}, \cr &&
\sinh^2(\la_1\rho)\cos^2(\la_2\te)={\rm e}^{2 z q_1^2}{\rm e}^{2 z
q_2^2}\bigl({\rm e}^{2 z q_3^2}-1 \bigr) ,\cr &&
\sinh^2(\la_1\rho)\sin^2(\la_2\te)\cos^2(\tes)={\rm e}^{2 z q_1^2}
\bigl({\rm e}^{2 z q_2^2}-1 \bigr),\label{xc} \\ &&
\sinh^2(\la_1\rho)\sin^2(\la_2\te)\sin^2(\tes)= {\rm e}^{2 z q_1^2}-1 ,
\nonumber
\eea where $z=\la_1^2$ and $\la_2\ne 0$ is an additional parameter
which can  be either a real  or a pure imaginary number~\cite{plb}.
Thus the metric (\ref{ccd}) is transformed into
\be
 \d s^2_{\rm I}=\frac {1}{\cosh(\la_1 \rr)}
 \left( \d \rr^2  +\la_2^2\,\frac{\sinh^2(\la_1 \rr)}{\la_1^2} \left( 
\d
 \te^2 + \frac{\sin^2(\la_2 \te)}{\la_2^2} \,\d\tes^2  \right) \right)
,
 \label{xd}
\ee which  is just the  metric of the 3D Riemannian and relativistic
spacetimes written in geodesic polar coordinates~\cite{Montreal}
multiplied by a global factor
${1}/{\cosh(\la_1 \rr)}\equiv {\rm e}^{-z J_-^{(3)}}$ that encodes the
information concerning the variable curvature of the space.  In the
new coordinates the sectional and scalar curvatures read
\be   K_{12}=K_{13}= -\frac 12 \la_1^2 \,\frac{\sinh^2(\la_1
\rr)}{\cosh(\la_1
 \rr)}, \qquad K_{23}=\frac{1}{2}K_{12},\qquad K= -\frac 52 \la_1^2
\,\frac{\sinh^2(\la_1 \rr)}{\cosh(\la_1
 \rr)} .
\ee

Therefore, by considering the possible pairs $(\la_1,\la_2)$ we have
obtained a quantum group deformation of the 3D sphere $(i,1)$,
hyperbolic   $(1,1)$, de Sitter   $(1,i)$ and anti-de Sitter $(i,i)$
spaces. The ``classical" limit 
$z\to 0$   corresponds to a   zero-curvature limit which leads to the
proper
  Euclidean $(0,1)$ and Minkowskian $(0,i)$ spaces. The resulting
integrable Hamiltonian on these six curved spaces with its two
constants of motion in the latter phase space turn out to be
\be {H}_z^{\rm I}=\frac 12 {\cosh(\la_1 \rr)}
 \left( p_\rr^2  + \frac{\la_1^2}{\la_2^2\sinh^2(\la_1 \rr)} \left(   
 p_\te^2 + \frac{\la_2^2}{\sin^2(\la_2 \te)}\,  p_\tes^2  \right)
\right) ,
 \label{ma}
\ee
\be {C}_z^{(2)}=p_\tes^2,\qquad {C}_z^{(3)}=p_\te^2+
\frac{\la_2^2}{\sin^2(\la_2 \te)}\,  p_\tes^2,
\label{mb}
\ee provided that
${H}_z^{\rm I}= 2{\cal H}_z^{\rm I}$, ${C}_z^{(2)}=4{\cal C}_z^{(2)}$
and
${C}_z^{(3)}= 4 \la_2^2 {\cal C}_z^{(3)}$.


\subsection{Superintegrable geodesic motion on spaces of constant
curvature}      

We  now consider the  Hamiltonian
${\cal H}_z^{\rm S}=\frac 12 \jp^{(3)}
\,{\rm e}^{ z \jm^{(3)}}$ which has four (functionally independent)
constants of motion, namely  ${\cal C}_z^{(2)}$ (\ref{bf}), ${\cal
C}_z^{(3)}$ (\ref{3cas}), ${\cal I}_z^{(2)}$ (\ref{bjj})
and~\cite{Deform} 
\be {\cal I}_z^{(3)}=\frac {\sinh z q_1^2}{2 z q_1^2} \, {\rm e}^{z
q_1^2} \, {\rm e}^{2 z q_2^2} p_1^2+\frac {\sinh z q_2^2}{2 z q_2^2}
\, {\rm e}^{z q_2^2}   p_2^2  .
\label{bbjj}
\ee Consequently ${\cal H}_z^{\rm S}$ determines a {\em maximally
superintegrable}  system with  free Lagrangian and  associated metric 
given, in terms of (\ref{ca}) and (\ref{ccd}), by
${\cal T}^{\rm S}_z = {\cal T}^{\rm I}_z \,{\rm e}^{-z \>q^2}$ and
$ \d s^2_{\rm S} = \d s^2_{\rm I}\, {\rm e}^{-z \>q^2}$. The 2D
pattern that links maximal  superintegrability and constant curvature
is reproduced in the 3D case since the space defined by $ \d s^2_{\rm
S} $ is of  Riemannian type with  constant sectional and scalar
curvatures   
$K_{ij}=z$,  $K=6z$. Next, through the change of coordinates
(\ref{xc}) and by introducing a new radial coordinate~\cite{plb}
$$r=\int_0^{\rr}\frac{\d x}{\cosh(\la_1 x)},$$ (i.e.~$\cosh(\la_1
\rr)=1/\cos(\la_1 r)$), we find that $ \d s^2_{\rm S} $ is just
transformed into a metric written in terms of   geodesic polar
coordinates~\cite{Montreal}
\be
 \d s^2_{\rm S}= \d r^2  +\la_2^2\,\frac{\sin^2(\la_1 r)}{\la_1^2}
\left(  \d
 \te^2 + \frac{\sin^2(\la_2 \te)}{\la_2^2} \,\d\tes^2  \right)   .
 \label{ye}
\ee According to the pair $(\la_1,\la_2)$, this metric  provides   the
usual (non-deformed) 3D sphere $(1,1)$, Euclidean $(0,1)$,
hyperbolic   $(i,1)$,  anti-de Sitter   $(1,i)$, Minkowskian $(0,i)$,
and de Sitter $(i,i)$ spaces.   In the latter phase space, the
maximal  superintegrable Hamiltonian, ${H}_z^{\rm S}=2{\cal H}_z^{\rm
S}$, is found to be
\be {H}_z^{\rm S}=\frac 12  
 \left( p_r^2  + \frac{\la_1^2}{\la_2^2\sin^2(\la_1 r)} \left(   
 p_\te^2 + \frac{\la_2^2}{\sin^2(\la_2 \te)}\,  p_\tes^2  \right)
\right) ,
 \label{yf}
\ee while its four constants of motion are the same ${C}_z^{(2)}$ and 
${C}_z^{(3)}$ (\ref{mb}), together with
\be {I}_z^{(2)}=\left(\la_2\sin(\la_2\te)\sin\tes\,
p_r+\frac{\la_1\cos(\la_2\te)\sin\tes}{\tan(\la_1 r)}\,p_\te + 
\frac{\la_1\la_2\cos \tes }{\tan(\la_1 r)\sin(\la_2\te)}\,p_\tes
\right)^2 ,
\ee
\be {I}_z^{(3)}=\left(\la_2\sin(\la_2\te) \,
p_r+\frac{\la_1\cos(\la_2\te) }{\tan(\la_1 r)}\,p_\te \right)^2 
\!\!\! + \la_1^2\la_2^2 \left(
 \frac{\tan^2(\la_1 r)\sin^2(\la_2\te)+1}{\tan^2(\la_1
r)\sin^2(\la_2\te)}\right) p_\tes^2,
\ee where ${I}_z^{(2)}=4\la_2^2{\cal I}_z^{(2)}$ and
${I}_z^{(3)}=4\la_2^2{\cal I}_z^{(3)}$. Notice that the two sets
$\{{H}_z^{\rm S}, {C}_z^{(2)}, {C}_z^{(3)}\}$ and $\{{H}_z^{\rm S},
{I}_z^{(2)}, {I}_z^{(3)}\}$ are formed by three functions which are
mutually in involution.

To conclude, the $N$D generalization of all of these results  
 should be obtained by following the same construction based on the
$N$-th coproduct map. In particular, the $N$D variable curvature spaces
will be defined through the 
$N$-particle symplectic realization of the free integrable
Hamiltonian~\cite{plb}
\be
 {\cal H}^{{\rm
 \SW},(N)}_z = \frac 12 
 \jp^{(N)}=\frac 12
 \sum_{i=1}^N
  \frac {\sinh z q_i^2}{z q_i^2}\,  p_i^2    \exp{\left( - z
 \sum_{k=1}^{i-1}  q_k^2+  z \sum_{l=i+1}^N   q_l^2 \right)} ,
 \label{ddr}
\ee and the $N$D analogue of the change of coordinates (\ref{xc}) will
be the cornerstone for the geometric interpretation of these spaces
and their free integrable systems. In the superintegrable case, ${\cal
H}_z^{{\rm S},(N)}=\frac 12 \jp^{(N)}
\,{\rm e}^{ z \jm^{(N)}}= {\cal H}_z^{{\rm I},(N)} {\rm e}^{ z
\jm^{(N)}}$, we conjecture that the $N$D analogues of the Riemannian
and relativistic spaces here presented will be recovered as the
corresponding constant curvature spaces defined by the maximal
superintegrable geodesic motion. Work on this line is in progress.

\bigskip

 \section*{Acknowledgements}  

{
 This work was partially supported  by the Ministerio de Educaci\'on y
Ciencia   (Spain, Project FIS2004-07913),  by the Junta de Castilla y
Le\'on   (Spain, Project  BU04/03), and by the INFN-CICyT
(Italy-Spain). }


\end {document}